\RequirePackage{snapshot}
\documentclass[aps,prx,twocolumn,showpacs]{revtex4-1}
\usepackage[dvipsnames]{xcolor}
\usepackage[colorlinks]{hyperref}

\hypersetup{
linkcolor=BrickRed
,citecolor=Green
,filecolor=Mulberry
,urlcolor=NavyBlue
,menucolor=BrickRed
,runcolor=Mulberry
,linkbordercolor=BrickRed
,citebordercolor=Green
,filebordercolor=Mulberry
,urlbordercolor=NavyBlue
,menubordercolor=BrickRed
,runbordercolor=Mulberry
}
\usepackage{graphics}
\usepackage{subfigure}
\usepackage{epstopdf}
\usepackage{epsfig}
\usepackage{placeins}
\usepackage{epsf,epic}
\usepackage{color}
\usepackage{marvosym }
\usepackage{amsmath}
\usepackage[utf8]{inputenc}
\usepackage{pgfplots}
\pgfplotsset{compat=newest}
\usepackage{amssymb}
\usepackage{amsfonts}
\usepackage{wrapfig}
\usepackage{multirow}
\usepackage{bm}
\usepackage{dcolumn}
\newcommand{\etal}{\textit{et al.\ }}
\newcommand{\ie}{\textit{i.e.\ }}

\begin{document}
\title{Distortion modes in halide perovskites: to twist or to stretch, a matter of tolerance and lone pairs}
\author{Santosh Kumar Radha$^a$, Churna Bhandari$^b$,  and Walter R. L. Lambrecht$^a$}
\affiliation{$^a$ Department of Physics, Case Western Reserve University, 10900 Euclid Avenue, Cleveland, OH-44106-7079}
\affiliation{$^b$ Department of Physics and Astronomy, University of Missouri, Columbia, Missouri 65211, USA}
\begin{abstract}
  Using first-principles calculations, we show that CsBX$_3$ halides
  with B=Sn or Pb undergo octahedral rotation distortions, while
  for B=Ge and Si, they undergo a ferro-electric rhombohedral distortion
  accompanied by a rhombohedral stretching of the lattice. We show that
  these are mutually exclusive at their equilibrium volume although
  different distortions may occur as function of lattice expansion.
  The choice between  the two distortion modes is in part governed by the Goldschmidt tolerance factor.
  However, another factor explaining the difference between  Sn and Pb compared with Ge and Si
  is the stronger lone-pair character of Ge and Si when forced to be divalent as is the case in these
  structures. The lone-pair chemistry is related to the off-centering. While the Si-based
  compounds have not yet been synthesized, the Ge compounds
  have been established experimentally. As a final test of the importance of
  the tolerance factor we consider RbGeX$_3$, which has 
  smaller tolerance factor than the corresponding CsGeX$_3$ because Rb is smaller than Cs.
  We find that it can  lower its energy by both rotations or rhombohedral
  off-centering distortions but the latter lower the energy slightly
  more efficiently. 
\end{abstract}
\maketitle
\section{Introduction}

The cubic perovskite structure is well known from the oxide perovksites to exhibit various possible phase transitions. These fall in two main categories: ferro-electric distortions, in which the $B$ atom in ABX$_3$ is displaced within its
surrounding octahedron, and antiferro-electric distortions, in which the octahedra rotate, possibly about multiple axes. Depending on the type of displacement,
for example along a cubic axis such as [001], or along two cubic axis or a
[110] direction, or three cubic axes, corresponding to the 
 [111] axis, the resulting symmetry becomes tetragonal, orthorhombic or
rhombohedral.  Likewise for the rotation type instabilities, rotation about one cubic axis leads to a tetragonal structure, about two orthogonal axes leads to
an orthorhombic phase.

The halide perovskites with B=Pb, Sn, Ge
have recently  garnered a lot of attention, mostly driven by the hybrid organic/inorganic halides' demonstrated potential  for solar cell applications.\cite{Kojima09,snaith01,snaith02,snaith03,snaith04,gratzel02,Kim14,Park15}
In particular methyl ammonium lead iodide (CH$_3$NH$_3$PbI$_3$ or (MA)PbI$_3$
or MAPI) and closely related materials have reached larger than 20 \% efficiencies in solar cells in a record development time frame.   The interplay between
the dipole character and the orientation of the organic component
and the inorganic framework
leads to interesting effects on the above mentioned phase transitions.\cite{Quarti14,Rinke16}
However, similar phase transitions also occur in the purely inorganic
CsBX$_3$ family.  While these distortions, which are related to soft-phonon
mode instabilities,\cite{Lingyi14}
lead to minor changes in the band structure, related to
bond angle distortions, other phases are known in the halides, which
are far more disruptive of the band structure. These latter phases include
edge-sharing octahedra and exhibit band structures with much wider band gaps
than their perovskite counterparts.\cite{Lingyi13}
As an example, the structural phases
in CsSnI$_3$ were studied in detail by In Chung \etal\cite{Inchungjacs}.
They fall generally in a set of three ``black phases'', cubic, tetragonal and
orthorhombic, which correspond to rotated octahedral structures,
and another orthorhombic ``yellow phase'', which has 1D chains of edge-sharing
octahedra forming Sn$_2$I$_6^{2-}$ structural motifs.  It is notable that
the transitions from cubic to tetragonal to orthorhombic perovksite each
time increase the density and the yellow phase has an even higher density.
The orthorhombic $\gamma$-phase is stable with respect to soft-phonons,
but has been calculated to have an energy either lower\cite{Lingyi16}
than or very close\cite{daSilva15} to that of the yellow phase.

Because the driving force for these transitions appears to be the
increasing density, the occurrence of the edge-sharing octahedral
structures, which appears to be detrimental for many of the sought applications,  may perhaps be already inferred from the behavior   of the material
under octahedral rotations, which in turn is related to the relative
sizes of the ions. For example, for the CsGeX$_3$ compounds, the sequence
of tetragonal, orthorhombic octahedral rotations is not observed and,
to the best of our knowledges, no edge-sharing octahedral phase is known
to occur, although a different, monoclinic phase occurs for the
Cl members of the family. 
Instead of octahedral rotation phases, a ferro-electric rhombohedral distortion is found to
occur in these materials, consisting of the displacement of the Ge
along the body diagonal of the cubic unit cell,
accompanied by a rhombohedral stretch of the unit cell.

In this paper
we examine the behavior of a family of halide perovskites computationally
under both
octahedral rotation and rhombohedral ferro-electric distortions.
Hence the phrase in the title: ``to rotate or to stretch''. We find that
the Sn and Pb members of the family of cubic perovskites are unstable toward 
rotation of the octahedra but stable with respect to ferro-electric
distortions. In contrast, the Ge and Si based compounds show the opposite behavior: they are unstable towards ferro-electric distortion but are stable with
respect to rotations. Furthermore
we relate this distinct behavior to the Goldschmidt tolerance factor,\cite{Goldschmidt1926} which
provides a convenient way to summarize the relative ionic sizes.
Notably, we include here the Si based halide perovskites, which have,
as far as we known, not yet been synthesized. 

The remainder of the paper is organized as follows. The details of our computational approach are given in Sec. \ref{method}. The relationships between the
different crystal structures and distortions to be studied are given in Sec. \ref{structures}. The results Sec. \ref{results}
is divided in several subsections. First,
we give a qualitative discussion in Sec. \ref{qualitative} establishing the
different behavior of Sn and Pb {\em vs.} Si and Ge. Next, we consider full
relaxations of the rotationally distorted structures of Sn and Pb based compounds in Sec. \ref{fullrelaxsnpb}, then the full relaxations of the rhomobohedral structures of the Ge and Si based compounds in Sec. \ref{fullrelaxgesi}. In Sec. \ref{latticeexp} we  study the competion between both types of distortion as function of lattice expansion for the Sn and Pb based systems. Finally
in Sec. \ref{Rbcase} we consider the RbGeX$_3$ compounds and end with a summary of the results in Sec.\ref{conclusions}.

\section{Computational Methods} \label{method}
The calculations are performed within density functional theory in the local
density (LDA) and/or generalized gradient (GGA) approximations.
Specifially, we use the Perdew-Burke-Ernzerhof (PBE) form of GGA.\cite{PBE}
The full-potential linearized muffin-tin orbital (FP-LMTO)
band-structure method is utilized.\cite{Methfessel,Kotani10}
Within this method, the basis set consists of Bloch sums of atom centered spherical waves as envelope functions,
described by smoothed Hankel functions,\cite{Bott98} which are then augmented with solutions of the radial Schr\"odinger equation inside muffin-tin spheres
and their energy derivatives. For the present calculations, a large basis set of $spdf-spd$ with two
sets of Hankel function decay constants $\kappa$ and smoothing radii is used. Inside the sphere,
augmentation is done to an angular momentum cut-off of $l_{max}=4$. The Cs $5p$ states are
treated as valence electrons. Likewise for Rb, the semicore $4p$ are treated
as local orbitals.
The Brillouin zone integrations are done with a $6\times6\times6$ $\Gamma$-centered
mesh. 

The LDA turns out to significantly underestimate the lattice constants in these materials, much more than the
GGA overestimates them. Although our initial study of the rotation or distortion patterns used the experimental lattice constants of the cubic phase,
our final full relaxation are done within GGA-PBE.

\section{Crystal structures}\label{structures}
\begin{figure}
  \includegraphics[scale=0.3]{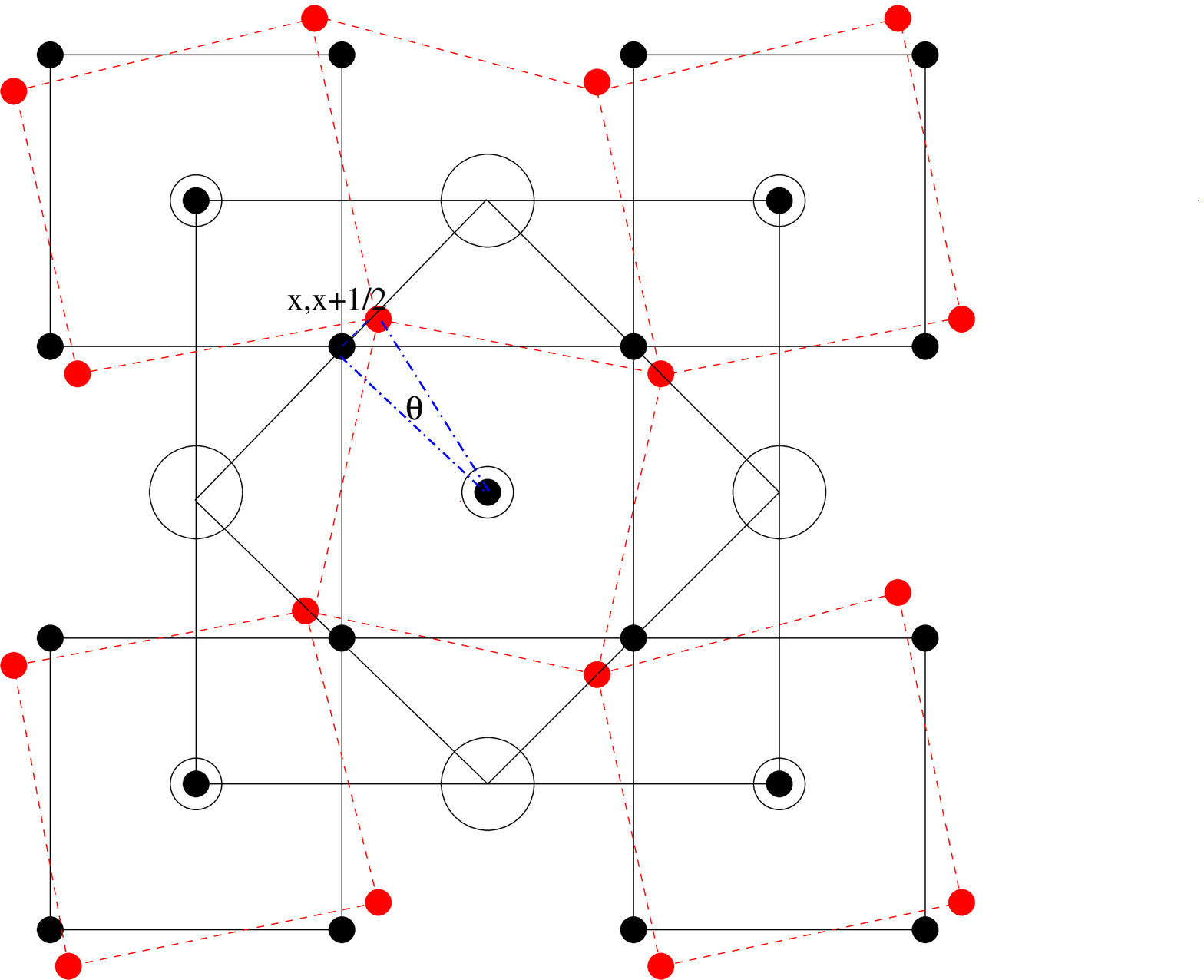}
  \caption{Rotation of octahedra in perovskite structure, large circles: A atom, small open circle: B atom,
    smallest filled circle: X atom, the black small circle corresponds to the cubic perovskite position,
    the red one the rotated one. The blue dash-dotted trangle indicates the rotation angle $\theta$.
    \label{figrotation}}
\end{figure}
We start from the cubic perovskite structure.  In this structure, with a simple cubic Bravais lattice,
for the composition ABX$_3$, the $B$ atom occurs in the center of the cubic unit cell and
is octahedrally surrounded by $X$ atoms on the face centers. The $A$ atoms occupy the corners of the cubic cell.
The stability is governed among other by the Goldschmidt tolerance factor, $t=(R_A+R_X)/\sqrt{2}(R_B+R_X)$,
where $R_A$, $R_B$, $R_X$ are the ionic radii.  Hence for $t=1$ the ionic spheres are touching and
hence Goldschmidt's original idea was that $t$ should not deviate too far from $1$ for the perovskite structure to
be stable. 
When $t<1$, the $A$ ion is somewhat too small for the interstitial space between the octahedra. This is what
leads to the rotations, which tighten the space for the $A$ ion. In contrast, when $t>1$, the octahedral
space is too large for the $B$ ion, which might then be expected to shift in its surroundings to make
stronger bonds with a subset of the 6 neighbors.  On the other hand,
it is not so clear {\em a-priori} whether this is related to the tolerance factor or to the
lone-pair character of the B-cation.

In terms of octahedral rotations, we consider both the in-phase and out-of-phase rotations about a single cubic axis.
These both lead to a tetragonal structure, the first one having the space group No. 127, $P4/mbm$ or $D_{4h}^5$,
the second one space group No. 140 , $I4/mcm$ or $D_{4h}^{18}$. They correspond to the Glazer tilt systems\cite{Glazer72,Woodward} $a^0a^0c^+$ and $a^0a^0c^-$ respectively.  Although other Glazer tilt systems are possible
and in fact occur in the Sn-halide perovskites,\cite{Lingyi14}
(a rotation about a second axis (Glazer $a^+b^-b^-$ leads to the orthorhombic $Pnma$  or $D_{2h}^{16}$ $\gamma$-phase),
we here are primarily concerned with
the instability either with respect to rotation of octahedra or ferro-electric distortions and thus
consider the tetrahedral rotation as the trigger toward rotation behavior.  So, we do not consider other
tilt systems. In the tetragonal $P4/mbm$ structure, the Wyckoff positions  for the B-atoms is 2a, for
the A-atom is 2c, for the X atoms, 2b and 4h. The $x$ parameter of the 4h positions is related
to the rotation angle of the octahedron by $\tan{\theta}=1-4x$ as can be seen in Fig. \ref{figrotation}.
In fact, the blue rectangular triangle marked by one corner at position $(x,x+\frac{1}{2})$ has sides
$(x-\frac{1}{4})\sqrt{2}$ and $\sqrt{2}/4$ in units of the lattice constant $a$
and hence their ratio gives $\tan{\theta}$.

As far as the ferro-electric distortions, we only consider the rhombohedral structure corresponding
to a displacement of the central $B$ ion along the [111]  direction. In the prototypical ferro-electric
oxide BaTiO$_3$ this phase occurs at the lowest temperatures, with an orthorhombic and tetragonal 
phase occuring at higher temperatures and eventually a cubic phase. Cooling from high temperature,
the displacement thus acquires successively more components along the cubic axes which deviate from the
central position. 
While we  presently do not exclude these other potential phases, our choice is guided by the CsGeX$_3$ compounds,
which have been found to exhibit this rhombohedral phase at low temperatures and a cubic phase at high
temperatures but no other phases in between.   The rhombohedral symmetry distortion of the ion is accompanied by a rhombohedral shear of the
lattice vectors.  Thus we will study the energy as function of displacement of the ion for varying rhombohedral
strain.

\begin{table}\begin{center}
    \caption{Shannon ionic radii ($R_i$) and tolerance factors ($t$) of cubic perovksites.
      The last column indicates whether the cubic structure is unstable toward octahedron rotation.
  \label{tabtolerance}}
\begin{ruledtabular}
\begin{tabular}{lcc}
Ion &$R_i$ (\AA) \\ \hline \\
Cs&     1.88  \\
Rb&     1.52 \\ \hline
Si&     0.4 \\ 
Ge& 0.53 \\
Sn& 0.69 \\ 
Pb& 0.775 \\ \hline
Cl& 1.81 \\
Br& 1.96 \\
I&  2.2  \\ \hline
Compound & $t$ & rotations\\ \hline \\
CsSiI$_3$ & 1.10 & no\\
CsGeI$_3$ & 1.057& no  \\
CsSnI$_3$ & 0.998& yes  \\
CsPbI$_3$ & 0.970& yes  \\ \hline
CsSiBr$_3$& 1.151& no \\
CsGeBr$_3$& 1.090& no \\
CsSnBr$_3$ & 1.025& yes \\
CsPbBr$_3$ & 0.993& yes  \\ \hline
CsSiCl$_3$ & 1.181& no \\
CsGeCl$_3$ & 1.115& no \\
CsSnCl$_2$ & 1.044& yes \\
CsPbCl$_3$ & 1.009& yes \\ \hline
RbGeCl$_3$ & 1.006&  yes  \\
RbGeBr$_3$ & 0.988&  yes  \\
RbGeI$_3$  & 0.964&  yes \\ 
\end{tabular}
\end{ruledtabular}
\end{center}
\end{table}

The occurrence of this distortion in Ge based halides but not in Sn or Pb based systems, which we will demonstrate
later, is not only related to the Goldschmidt ratio of ionic sizes but is also
related to the lone-pair character of the bonding. 
As one goes down the
column of group-IV atoms, the valence $s$ states become increasingly deeper relative to the valence $p$
states. That is why carbon  has $s$ and $p$ orbitals of similar extent and is extremely flexible in choosing
different hybridization schemes: $sp^2$ in graphite, $sp^3$ in diamond and so on. Si and Ge clearly prefer
$sp^3$ hybridization and thus tend to be tetravalent, while Sn and Pb become increasingly divalent.
Nonetheless, in the halide perovskite crystal structure,  it is clear that even Ge behaves as a divalent
ion. 
Whether Si can also be forced to be divalent in these compounds remains to be seen. 
However, the $s$-electrons then behave as a stereochemically  active lone-pair, which
promotes off-centering of the Ge in its surrounding octahedron with an asymmetric bonding configuration in which the lone
pair electrons are located opposite to the direction of the displacement of the ion.\cite{Walsh11}
The lone-pair related trends in the series Pb-Sn-Ge have been addressed by Waghmare \etal\cite{Waghmare03} in the context
of IV-VI compounds. 
We will show that even in the Sn-case this happens under lattice expansion, as was previously shown
by Fabini \etal\cite{Fabini16} According to the latest insights into lone-pair chemistry, the hybridization
with the anion $p$-orbitals play a crucial role in this. The important
role of the Sn-$s$ halogen-$p$ hybridization on the band structure of CsSnX$_3$ halides was already
pointed out in our earlier work.\cite{Lingyi13} We point out here that competition between rotation instabilities and
lone-pair off-centering was previously studied in CsPbF$_3$ by Smith \etal\cite{Smith15}
Lone pair physics related to Pb also occurs when Pb is the A-cation in oxide perovskites.\cite{Waghmare97,Ghosez99}

Finally, we should mention that the tolerance factor depends on the choice of ionic radii. Usually
we use the Shannon\cite{Shannon76} ionic radii for this purpose. However, these are themselves based on an analysis
of bonding in different coordinations and for example do not give us any information on the behavior
under hydrostatic pressure. One might conceivably think of the relative ion sizes to change with
pressure or wish to include other aspects than pure ionic size to predict structural stability.\cite{Brehm14,Travis16}
With these precautions, we used the Shannon ionic radii calculated tolerance factors as a guide to our study.
They are summarized in Table \ref{tabtolerance}. We note that our goal with the tolerance factor is not
so much to predict structural maps in the sense of separating perovskite versus non-perovskite forming compounds
but rather the type of structural distortion occurring within the perovskite. Also, because Shannon only
provides ionic radii for Pb(II) in the divalent state, but not for Sn, Ge or Si, we used instead
the tetravalent radii for octahedral environment.  This may seem to contradict the fact that in these
structures the B ion is supposed to be divalent. On the other hand, we should recognize that the bonding
is partially covalent anyways.   We find that within each group of a given anion, the tolerance
factor decreases along the sequence  Si-Ge-Sn-Pb.
The dividing critical value between octahedral rotations being
favored or not, depends actually on which anion (a similar point was
also made by Travis \etal\cite{Travis16}),
but is close to 1 for all the Cs compounds. For the Cl compounds, it would be between 1.115 and 1.044. 
For Rb which has a smaller radius, the value 1.006 is thus definitely on the small side and hence
predicts rotations to occur.  

Our goal in this paper is to study the instability of the cubic structure to these two types of
distortion as function of the $B$ atom and to correlate them with the tolerance factors in
the above Table \ref{tabtolerance}.

\section{Results} \label{results}
\subsection{Qualitative discussion}\label{qualitative}
\begin{figure}
\includegraphics[scale=0.35]{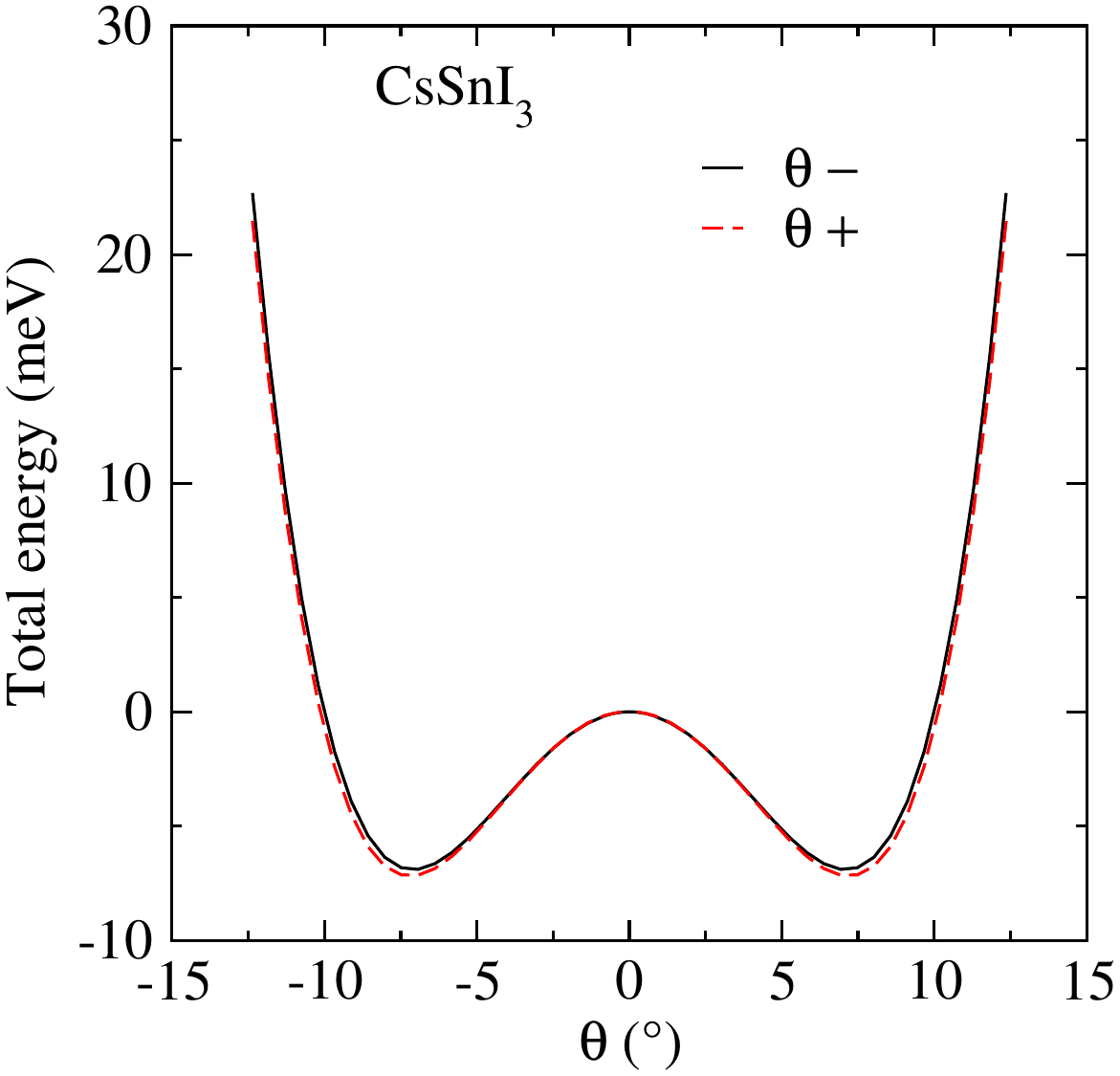} \includegraphics[scale=0.35]{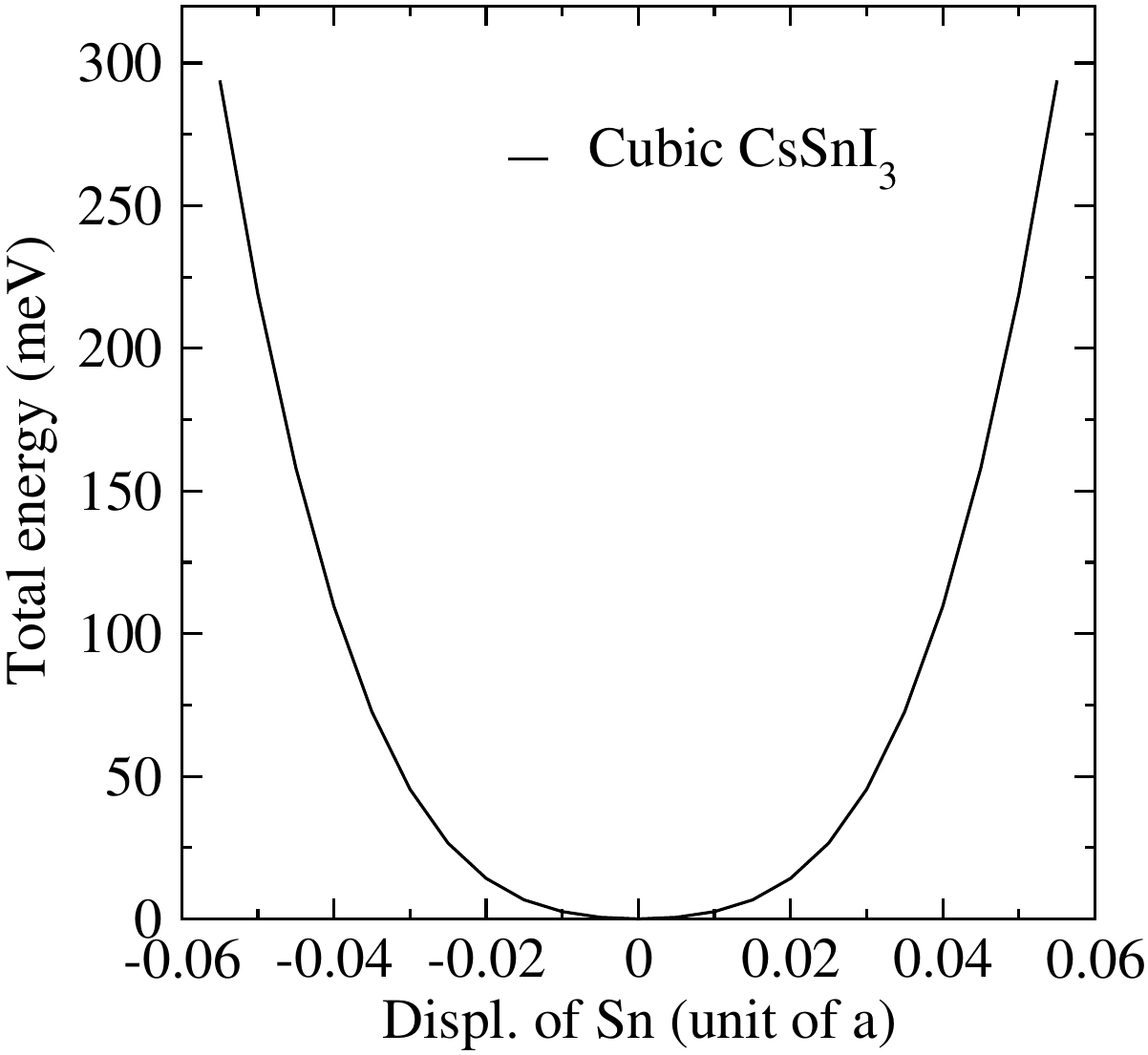}
\caption{left: Total energy (per formula unit in meV) vs. octahedral rotation angle $\theta$ ($^\circ$) in CsSnI$_3$ left. Here $\theta +$  stands for out-of-phase rotation and $\theta -$ for in-phase rotation respectively. Right: total energy vs. displacement of Sn from body center in unit of the cubic lattice constant $a$.\label{csi1}}
\end{figure}

First, we consider the CsSnI$_3$ compound. In Fig. \ref{csi1}a we show its total energy as function
of rotation angle $\theta$ of the octahedra. This calculation is done at
the cubic experimental volume
although we know that the observed $\beta$-structure, corresponding to the $P4/mbm$ space group
has higher density. We consider both the in-phase and out-of-phase rotations. The figure shows that their
energy is almost indistinguishable. More importantly, it shows clearly that the system prefers a rotation
angle of about 6.9$^\circ$. Of course, the rotation can be either clockwise or counterclockwise.  The energy
barrier between the two is of the order of a few meV/formula unit.
So, this agrees with the well-established fact that CsSnI$_3$ undergoes octahedral rotations of this type
although the equilibrium  optimum angle appears to be somewhat underrestimated compared
to the experimental angle which is 9$^\circ$, corresponding to the Wyckoff parameter $x=0.21$.
This is because in this initial calculation, we kept the cubic structure of the lattice and
did not allow yet for a full relaxaton. Full relaxation results are given later in Table \ref{tabrot} and are discussed in subsection \ref{fullrelaxsnpb}.
The present result shows that the
rotation instability is already present even at the volume of the cubic structure.

Next, we consider the behavior of CsSnI$_3$ under the ferro-electric rhombohedral distortion.
We do this at zero strain, so keeping the cubic lattice vectors. Clearly there is only one minimum
at exactly the central position of the Sn in  its octahedral cage. So, there is no evidence for
a ferro-electric  instability. Nonetheless, the curves are clearly not
parabolic but show a rather flat energy minimum region for the position
of the central atom. 

\begin{figure}
\includegraphics[scale=0.35]{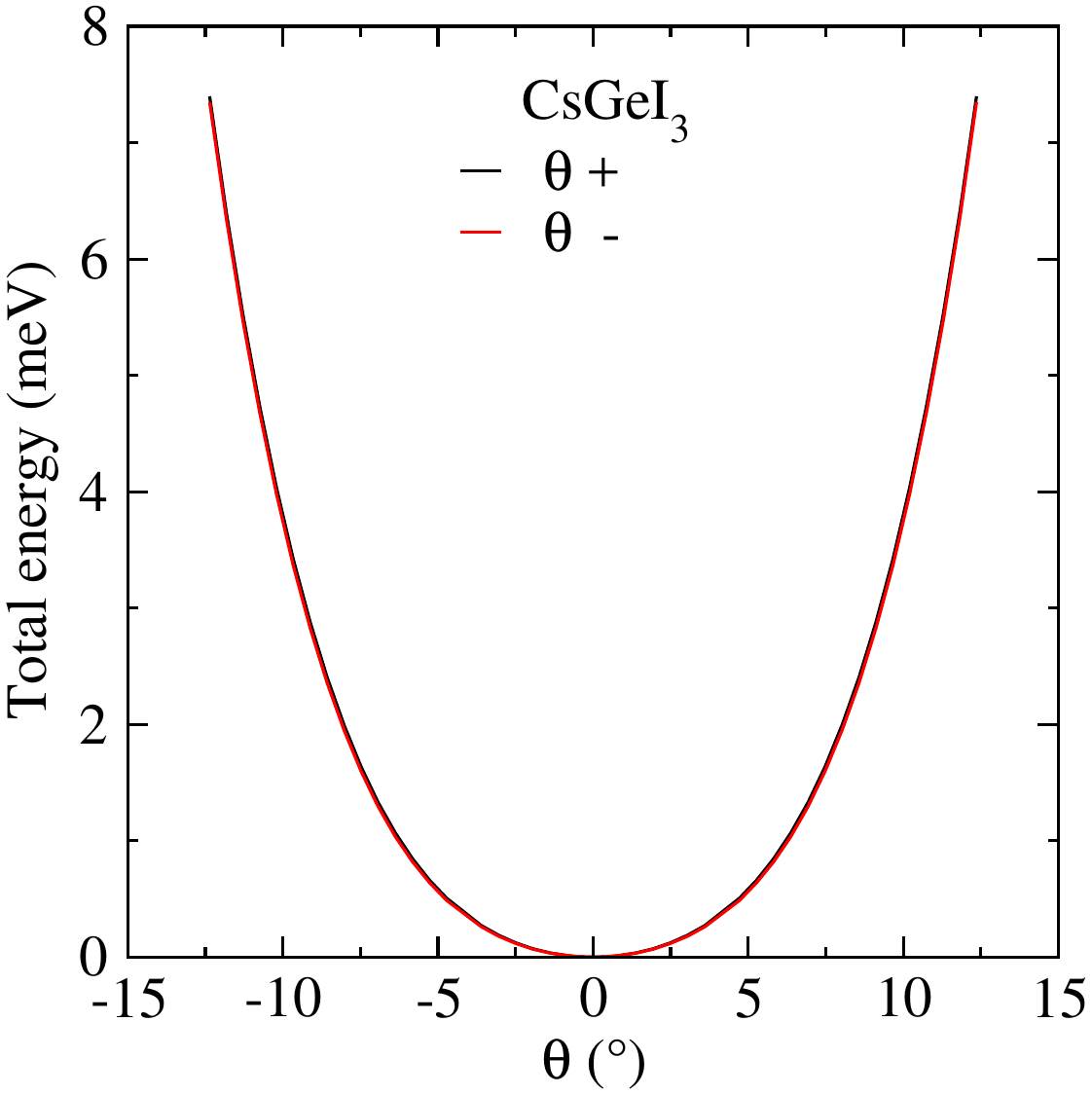}
\includegraphics[scale=0.35]{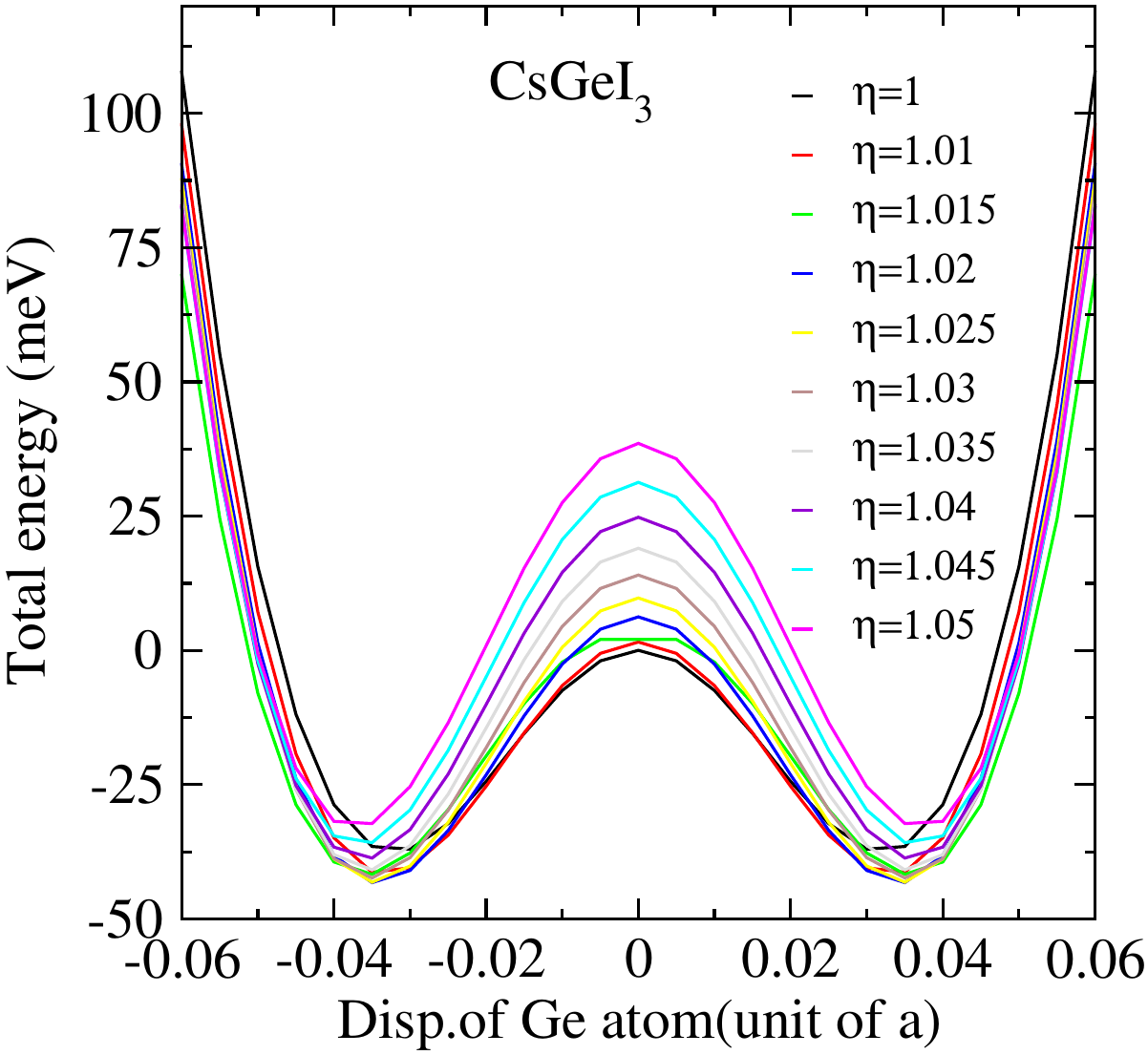}
\caption{Left: Total energy vs. octahedral rotation angle $\theta$ ($^\circ$) in CsGeI$_3$ left. Here $\theta +$  stands for
  out-of-phase rotation and $\theta -$ for in-phase rotation respectively. Right: total energy vs. displacement of Ge from body center in units of the lattice constant $a$.\label{csge1}}
\end{figure}

In contrast, if we consider CsGeI$_3$, (Fig. \ref{csge1}a) for its rotational stability, we find no evidence
at all of a rotational instability. The preferred angle is 0. This is true for both in-phase and out-of-phase
rotations.  On the other hand, in Fig.\ref{csge1}b we see that now there is a clear instability
against the ferro-electric displacement. Again, it is symmetric with respect to the central position.
The displacement is given in units of the lattice constant $a$ of the cubic cell. The optimimum
position lies between 0.52 and 0.54 or 0.46 and 0.48. In this case, we study the optimum
position and the energy barrier as function of rhombohedral strain but initially keeping the volume fixed
at that of the cubic structure. This is quantified by the parameter
$\eta$ which gives the stretch  along the [111] direction (when $\eta>1$) and is compensated by a compression
in the orthogonal directions, which conserves the volume. Thus, we applied here a pure shear or traceless strain at
fixed volume. We can see that the optimum position varies slightly with the strain. The lowest overall energy
occurs for a strain of $\eta=1.03$ and $\delta u=0.035$. 
A full structural relaxation within the rhombohedral symmetry requires not only optimizing $u$ and $\eta$
but also the volume and the results of such a full relaxation are given in Table \ref{tabrhom} in Sec. \ref{fullrelaxgesi}.

We thus see a mutually exclusive behavior of the two type of distortion modes. Either the material
is unstable under rotations, or it is unstable under the ferro-electric distortion but not both.
We found that these structural instabilities already occur at the cubic structure equilibrium
volume but once the distortion takes place and full relaxation is allowed, a new equilibrium is found.
We should remember though that the mutual exclusivity
correspond to the experimental volume. This might change
as function of pressure. For example, in SrTiO$_3$, Zhong and Vanderbilt\cite{Zhong95} predict an interplay
between the two types of distortions, leading eventually to a complex phase diagram as function
of pressure and temperature. We will discuss the distortion behavior for CsSnI$_3$ as function of
lattice constant later. 

Having established the basic two types of behavior, we now consider the variation with anion.
In the CsSnBr$_3$ and CsSnCl$_3$ cases, we again find the structure to be stable against ferro-electric
distortion, but unstable toward rotations. The energies as function of rotation angle are given in Suppelementary Material.\cite{supinfo}
For the CsGeBr$_3$, CsGeCl$_3$ cases, we find the structures to be stable under rotation as expected but
we do find a ferro-electric distortion in both cases.\cite{supinfo}
Next we show that Pb behaves similar to Sn and Si behaves similar to Ge.\cite{supinfo}
For the Si case, where no experimental results are known, we initially
used the LDA optimized lattice constants for the cubic CsSiX$_3$ case
but in the next section for our fully relaxed structures, we use GGA-PBE
for improved accuracy.

\subsection{Full structural relaxation for Sn and Pb based rotations.}\label{fullrelaxsnpb}
\begin{figure}[ht]
 
 \includegraphics[width=9cm,keepaspectratio]{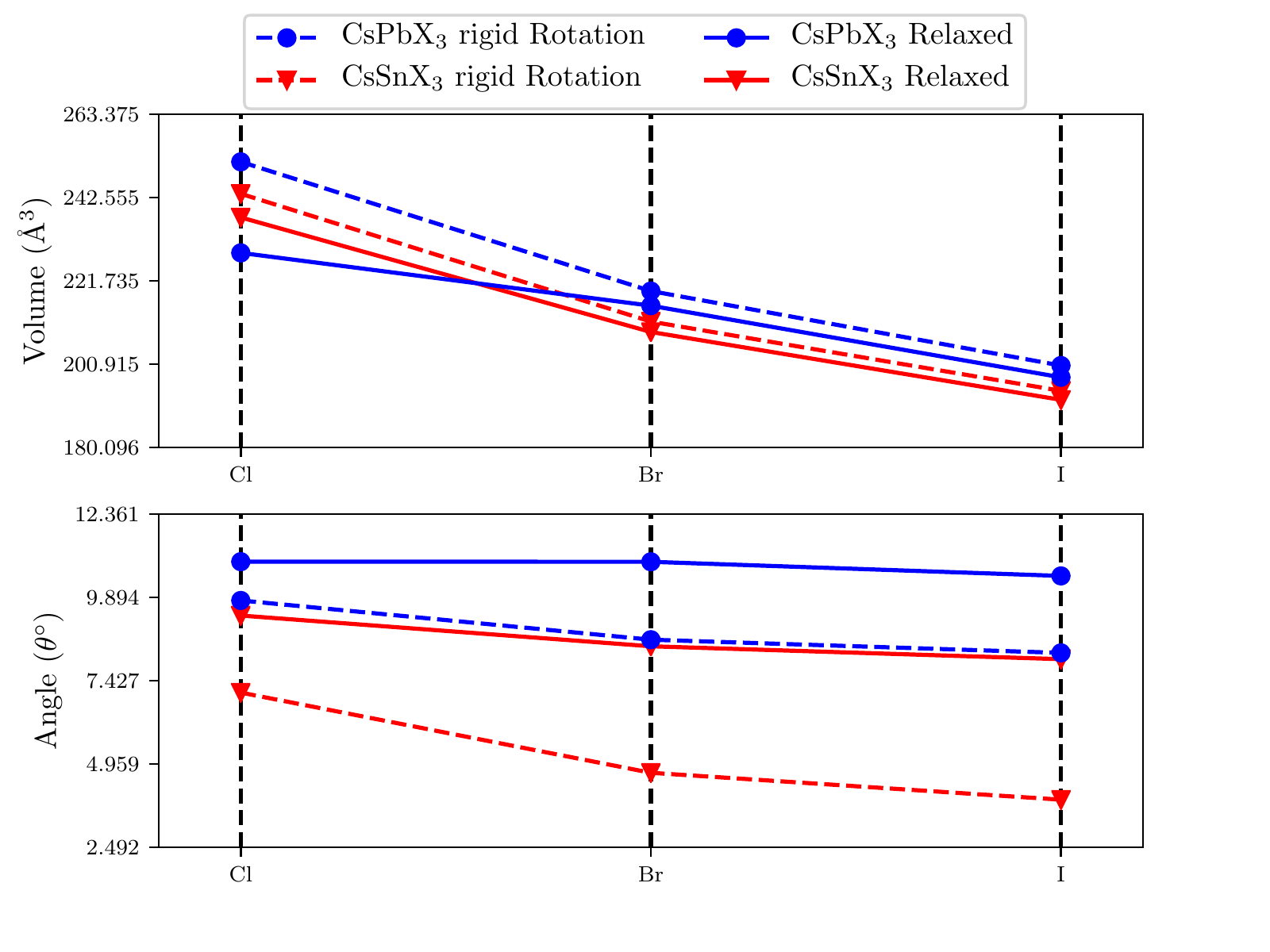}

\caption{Optimized volume and rotation angle as function of halogen in CsSnX$_3$ and CsPbX$_3$.\label{figoptrot}}
\end{figure}

In this section we study the fully relaxed tetragonal $P4/mbm$ structure corresponding to the
rotational distortions. 
The optimum rotation angles are summarized in Table
\ref{tabrot}.  Because we found LDA to underestimate the lattice constants more than GGA overstimates
them, we performed the full structural relaxations in GGA-PBE.  In Table \ref{tabrot}
we show both the results for the rotation angle when fixing the lattice constants to be ``rotated cubic''
and fully relaxing the tetragonal structure, \ie also  relaxing $c/a$.  By  ``rotated cubic'' we mean we consider a
$\sqrt{2}\times\sqrt{2}$ superlattice in which the octahedrons can rotate as shown in Fig. \ref{figrotation}
but keep the $c/a$ ratio exactly at a factor $\sqrt{2}$ and keep the volume at the cubic volume.  These results are also  presented
in Fig. \ref{figoptrot} to visualize the trends with halogen.
\\
\begin{table}
  \caption{\label{tabrot} Structural relaxation results for rotation for the CsSnX$_3$ and CsPbX$_3$
    compounds: $\alpha^\prime$ means ``rotated cubic'' and $\beta$ means fully relaxed tetragonal. All results
    obtained within GGA-PBE. Volume is per formula unit. $\Delta E$ is
    the energy barrier between the optimum angle structure and
  the cubic structure at rotation angle $\theta=0$.}
  \begin{ruledtabular}
    \begin{tabular}{l|cc|cc|cc}
      Compound & \multicolumn{2}{c}{CsSnI$_3$} & \multicolumn{2}{c}{CsSnBr$_3$} & \multicolumn{2}{c}{CsSnCl$_3$} \\ 
      Structure& $\alpha^\prime$& $\beta$ &  $\alpha^\prime$& $\beta$ &    $\alpha^\prime$& $\beta$  \\ \hline \\
      $a$ (\AA) & 8.935 &  8.800        & 8.372  &  8.282  &  8.033  &  7.942 \\ 
      $c$ (\AA) & 6.318 &  6.300        & 5.920  &  5.944  &  5.78   &  5.710 \\
      $V$  (\AA$^3$) & 252.19 & 243.92   & 207.47 & 203.84  &  183.25 &  180.10 \\
      $\Delta V/V$ (\%)      &  & -3.28 & &-1.75           & & -1.72   \\
      $\theta$ ($^\circ$)  & 6.93 & 10.1 & 3.61   & 8.85   &  2.49    &  8.32 \\  
     $\Delta E$ (meV) &  9.9 & 11.2 &  4.6   & 6.6    &  0.7  & 8.9  \\
       \hline 
     Compound & \multicolumn{2}{c}{CsPbI$_3$} & \multicolumn{2}{c}{CsPbBr$_3$} & \multicolumn{2}{c}{CsPbCl$_3$} \\ 
     Structure& $\alpha^\prime$& $\beta$ & $\alpha^\prime$& $\beta$ &  $\alpha^\prime$& $\beta$  \\ \hline \\
     $a$ (\AA) &   9.065 & 8.610           &  8.514 & 8.367    &  8.160  &  8.034 \\
     $c$ (\AA) &   6.410 & 6.245           &  6.020 & 6.085    &  5.77   &  5.82  \\
     $V$ (\AA$^3$) & 263.37 & 231.520      &  218.17& 213.039  &  192.10 &  187.98 \\
     $\Delta V/V$ (\%) & & -12.09          &        & -2.35    &         &  -2.15  \\
     $\theta$ ($^\circ$) &  10.75 & 12.36 &  9.13   & 12.36    &  8.58   & 11.77   \\
     $\Delta E$ (meV) &  33 & 258    & 22    & 50 & 17   & 39  
    \end{tabular}
\end{ruledtabular}
\end{table}

For CsSnI$_3$, there are two sets of experimental results, by Yamada \etal\cite{Yamada91} and by In Chung \etal\cite{Inchungjacs}.
Yamada \etal give $a=8.772$ \AA, $c=6.261$ \AA, $V=240.815$ \AA$^3$ for the $\beta$-structure and
$a=6.219$ \AA, $V=240.526$ \AA$^3$ for the cubic structure, in other words, almost the same volume.
In contrast, In Chun \etal\cite{Inchungjacs} give $a=8.7182$ \AA, $c=6.1908$ \AA, $V=235.27$ \AA$^3$ for the tetragonal and
and $a=6.2057$ \AA, $V=238.99$ \AA$^3$ for the cubic structure. These results correspond to 500 and 380 K respectively
and clearly show a smaller volume for the tetragonal structure. Our calculated results agrees qualitatively better with
those of In Chun \etal\cite{Inchungjacs} in finding a volume reduction induced by the $\alpha\rightarrow\beta$
transition. Our GGA calculations overestimate
the experimental volumes by about 5.6\% and 3.6 \%  for the cubic and tetragonal structures compared to In Chung \etal\cite{Inchungjacs}
We find systematically the same trend in volumes for the other compounds.
We may note that the optimum rotation angle depends strongly on volume. It is typically larger in the relaxed
tetragonal $\beta$-structure than if we keep the volume fixed at the cubic volume.  We may also note that it decreases
with decreasing volume along the series CsSnI$_3$, CsSnBr$_3$, CsSnCl$_3$  and similarly in the Pb based series.
The rotation angles are larger in the Pb-based compounds than in the Sn-based compounds. This means the smaller the
tolerance factor, the larger the rotation.  Our optimum angle of octahedral rotation for CsSnI$_3$ agrees well with the
experimental value of 9.09$^\circ$.\cite{Yamada91}

Finally we may note that for the larger cubic volumes, the rotation angle for the Sn-based compounds becomes rather small.
Below, in Sec. \ref{latticeexp}, we show that under lattice constant expansion it actually goes to zero, and, at some critical volume, the
ferro-electric distortion becomes preferable instead. 

The energy barriers $\Delta E$ between the tetragonal energy minimum and the
cubic unrotated structure are seen to be significantly larger for the
Pb compounds than the Sn compounds and within each family decrease from I to Br to Cl, except for the fully relaxed CsSnBr$_3$ and CsSnCl$_3$.

\begin{figure*}[ht]
\includegraphics[width=.99\textwidth,keepaspectratio]{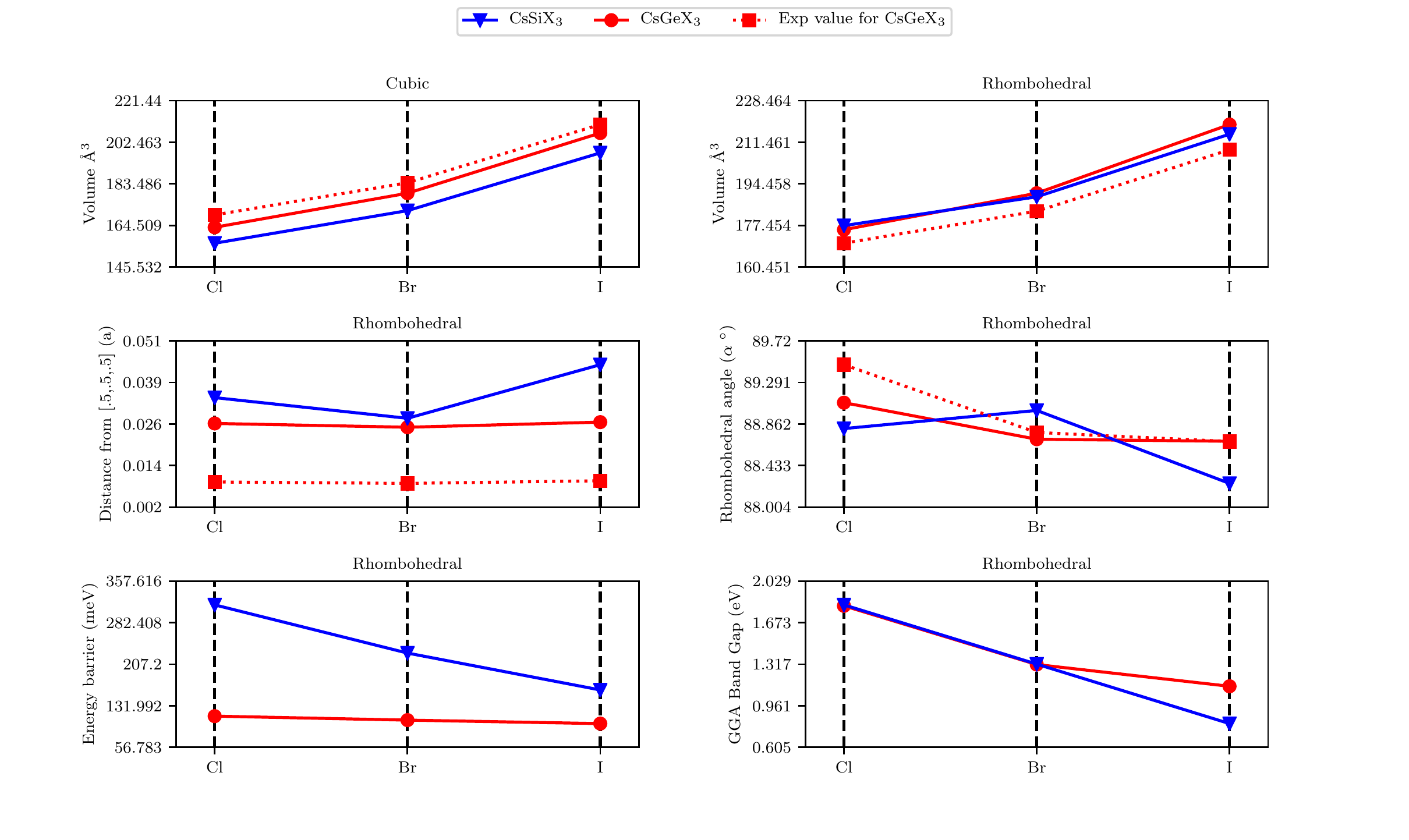}

\caption{Trends in structural relaxation parameters for the CsGeX$_3$ and CsSiX$_3$
  halides corresponding to the data in Table  \ref{tabrhom}\label{figdistort}}
\end{figure*}

\subsection{Full structural relaxation for Ge and Si based rhombohedral distortions.}\label{fullrelaxgesi}

\begin{table}
  \caption{Optimized cubic and
    rhomobohedral structures for CsBX$_3$ with B=Ge, Si.
    The $\Delta E$ are the barriers between the cubic structure
    with $\delta u=0$ and $\eta=1$ and the optimized rhombohedral structure
    each at their own equilibrim volume.
    \label{tabrhom}}
  \begin{ruledtabular}
    \begin{tabular}{lccc}
      Compound & CsGeCl$_3$ & CsGeBr$_3$ &  CsGeI$_3$ \\ \hline \\
      cubic $V$ GGA (\AA$^3$)    & 155.72 & 177.50 & 216.00 \\ 
      cubic $V$ Expt.(\AA$^3$) & 163.67 &  184.22 & 221.44 \\
      rhombohedral $V$  (\AA$^3$) & 168.19 & 189.11 & 228.46 \\
      rhombohedral $a$ (\AA) GGA    & 5.52  & 5.74  & 6.11 \\
      rhombohedral $a$ (\AA) Expt   & 5.434  & 5.635 & 5.983  \\
      $\delta u$             & 0.027  & 0.026  & 0.028 \\
      $\delta_{1}$              & 0.015    & 0.011    &0.008     \\
      $\delta_{2}$              & 0.022    & 0.013    &0.004     \\
      $\eta$                 & 1.014  & 1.023  & 1.024 \\
      $\alpha$  GGA          & 89.17  & 88.64  & 88.61 \\
      $\alpha$  Expt.        & 89.72  & 88.74  & 88.61 \\
      $\Delta E$ (meV)       &  75      & 65   & 56      \\
      $E_g$ (eV) GGA         & 2.01   & 1.31   & 1.05   \\
      \hline \\
      Compound               & CsSiCl$_3$ & CsSiBr$_3$ & CsSiI$_3$ \\ \hline \\
      cubic $V$              & 145.531     & 166.375 & 203.297 \\
      rhombohedral $V$ (\AA$^3$) & 170.53 & 187.06 & 222.91 \\
      rhombohedral $a$ (\AA)     & 5.54   & 5.71   & 6.06 \\
      $\delta u$                 & 0.038   & 0.029   & 0.052 \\
      $\delta_{1}$              & 0.007    & 0.010    &0.007    \\
      $\delta_{2}$              & -0.057    & 0.033    &0.018     \\
      $\eta$                     & 1.021 & 1.016   & 1.034 \\
      $\alpha$                   & 88.79   & 89.05   & 88.00  \\
      $\Delta E$ (meV)           & 357     & 235     & 142    \\
      $E_g$ (eV) GGA             & 2.02    & 1.31    &0.605     \\
      
      \end{tabular}
  \end{ruledtabular}
\end{table}

In this section we further study the fully relaxed rhombohedrally distorted structures. 
In Table \ref{tabrhom} we first give the optimum GGA volume of the cubic
structure. It is compared with the experimental values
at elevated temperature where that phase is stable, from Thiele \etal\cite{Thiele87} at 170, 270 and 300 $^\circ$C respectively for the Cl, Br, I cases. Clearly these values are larger than our GGA because of the lattice expansion at elevated temperature.
Next, we applied a rhombohedral strain along the cubic structure, allowed the  central Ge atom to
go off-center by a displacement $\delta u$ and
allowed the volume to relax. The strain $\eta=1+2\epsilon$ 
is applied along the [111] cubic direction while perpendicular to it,
the distances are multiplied by $1/\sqrt{\eta}\approx 1-\epsilon$, thus
maintaining the volume. The strain tensor can be written to linear order
\[ \bm{\epsilon}= \left ( \begin{array}{ccc}0 & \epsilon& \epsilon \\
  \epsilon& 0 & \epsilon \\ \epsilon & \epsilon & 0 
\end{array} \right ) \]
The cubic lattice vectors $a[1,0,0]$ are thus distorted
into vectors of $a(1,2\epsilon,2\epsilon)$ with length $a\sqrt{1+2\epsilon^2}$
which to first order in $\epsilon$ means they stay unchanged.
The results for the Ge and Si based compounds are given in Table \ref{tabrhom}.
We can see that for the Br and I cases, our relaxed lattice constant
for the rhombohedral phase in GGA
slightly overestimates the experimental value, even though the latter
is measured at 20$^\circ$C while our calculated volume is in principle at
0 K. For the Cl case the calculated lattice constant is slightly underestimated. 

The displacement from the 0.5 value is almost the same in all cases.
The rhombohedral angle extracted from the shear $\eta$ using
$\cos{\alpha}=\eta_{min}-1$ to linear order in strain agrees well with the 
experimental values. 
For the Si compounds, all values are obtained within GGA and no experimental
values are available to compare with.

The full relaxation also requires the anions to move. For example the
anion which in the cubic case is located at (0.5, 0, 0.5) moves to
$(0.5-\delta_1,-\delta_2,0.5+\delta_1)$, in other words, it moves inward toward the
displaced Ge as shown in Fig.\ref{figrotation2}. The motion of the other anions is similarly determined by
symmetry. The corresponding parameters are given in Table \ref{tabrhom}.
Table \ref{tab:bondlength} shows that the B-X bond lengths are shortened
upon relaxation in spite of the overall volume being expanded in the
rhombohedral distortion. 

\begin{table}[]
\centering
\caption{B-X bond length ( in \AA) compared between the perfect cubic structure and the relaxed structure, where B={Ge,Si} and X={Cl,Br,I}}
\label{tab:bondlength}
\begin{ruledtabular}
\begin{tabular}{lccc}

Compound & cubic & relaxed & \% change \\ \hline \\
CsGeCl3  & 2.69  & 2.49    & -7.88\%   \\
CsGeBr3  & 2.81  & 2.65    & -5.93\%   \\
CsGeI3   & 3     & 2.86    & -4.77\%   \\
CsSiCl3  & 2.63  & 2.31    & -13.91\%  \\
CsSiBr3  & 2.75  & 2.50    & -9.80\%   \\
CsSiI3   & 2.94  & 2.68    & -9.62\%   \\ 
\end{tabular}
\end{ruledtabular}
\end{table}

The energy differences $\Delta E$ between the cubic undistorted
structure and the rhombohedral optimized structure each at their own
equilibrium volume are also shown in Table \ref{tabrhom}.
They indicate an increase from Cl to Br to I and much larger values for the
Si then the Ge based compounds. 
\begin{figure}
\includegraphics[scale=.11]{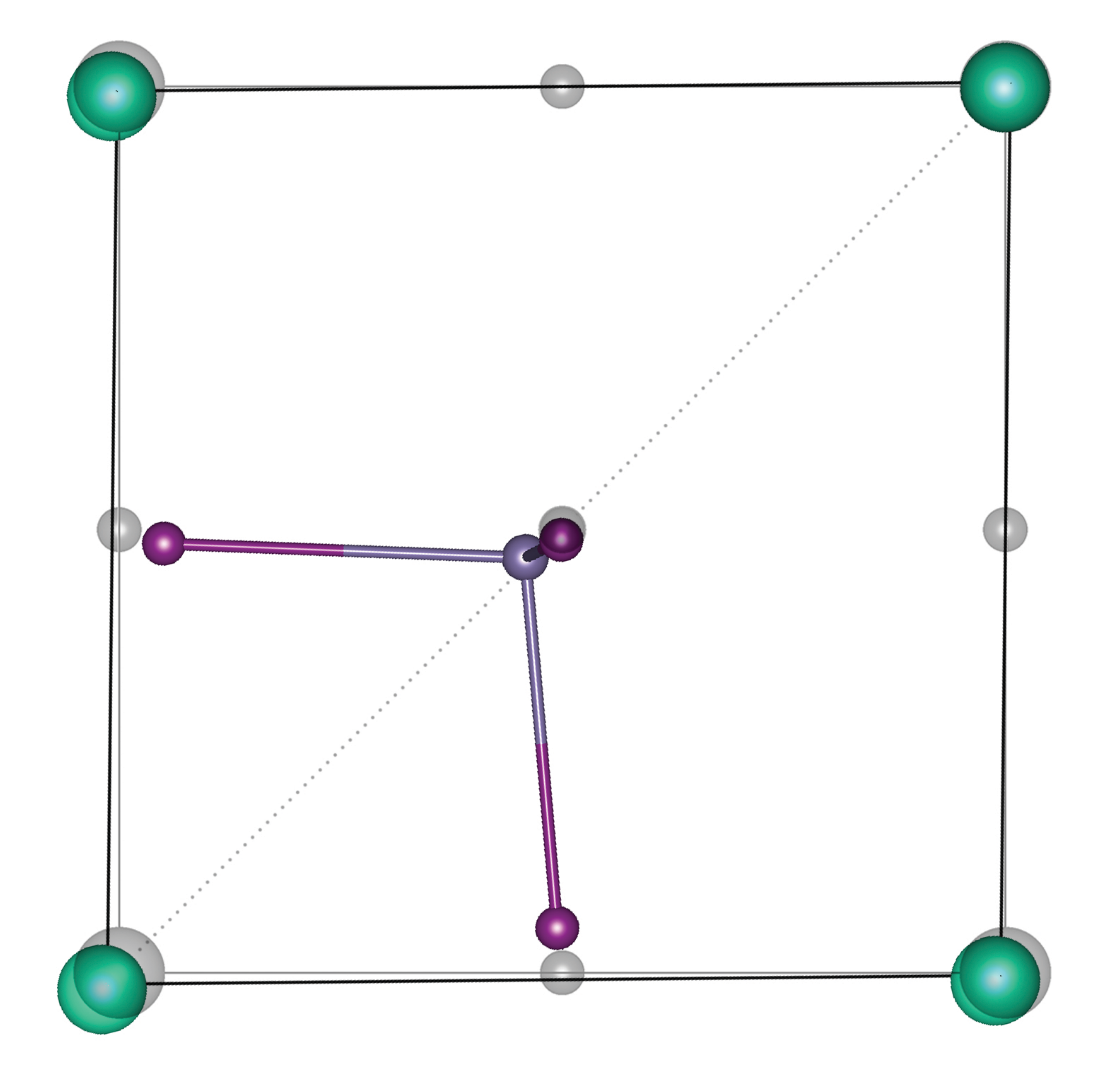}

\caption{Unit cell of the relaxed structure of CsSiCl$_3$ with the colored atoms at the relaxed positions and the gray atoms at the unrelaxed cubic positions. Green pink and violet spheres represent Cs, Cl and Si respectively. \label{figrotation2}}
\end{figure}

The band gaps, which must be underestimates because of the GGA, are also
included in Table \ref{tabrhom}
and show the expected trend of decreasing from Cl to Br to I,
in other words decreasing with decreasing ionicity. They are also smaller in the Si than the
Ge compounds. 
The  gaps in the $GW$ approximation
at the experimental rhombohedral structures for the CsGeX$_3$ compounds
were given in Ref. \onlinecite{Lingyi16} and are
4.304, 2.654 and 1.619 eV for the Cl, Br and I cases respectively. 
For the Si- based compounds, they remain to be determined but assuming
a similar gap correction, we can already see that both CsSiI$_3$ and CsSiBr$_3$
may have gaps suitable for photovoltaics. 
The trends of the data in Table \ref{tabrhom} are visualized in Fig. \ref{figdistort}.

Although the energy barriers increase from Cl to Br to I, they do not
show a clear correlation with the transition temperatures, which
are 277-283 $^\circ$C, 238-242 $^\circ$C, and 155 $^\circ$C
respectively for CsGeI$_3$, CsGeBr$_3$, CsGeCl$_3$. 
The problem here is that our calculations consider a homogeneous
transformaton, which is forced to be the same in each unit cell. In the actual phase transition,
there is a competition between the  interaction energies of atoms 
in neighboring cells and the double-well anharmonic
potential well in each unit cell. The phase transition could be either displacive
or order-disorder type.\cite{Dove} In the former case, corresponding to a large interaction
energy between neighboring cells, the positions of the atoms vibrate about an average
near the barrier maximum (corresponding to the cubic structure)
at high temperature and settle into one or the other minimum
below the transition temperature. A nucleation process occurs where groups of neighboring atoms
settle into one of the two local minima. In contrast in the order-disorder model, corresponding
to a strong double well potential but weaker intercellular interactions, 
the atoms are always in one of the two minima but at high temperature, they are equally
likely to be in the left or right well. From our present calculations, we do not
have access to the inter-cell energies in such a model,
and thus we cannot draw conclusions  about the
nature of the phase transition. Experimentally, it was established by Thiele \etal\cite{Thiele87}
that for CsGeBr$_3$ and CsGeI$_3$ the phase transition is first-order, while for
the Cl-case it is second order. This would indicate a displacive transition for the latter case
but an order-disorder type for the former.

\subsection{Rotation and rhombohedral distortion under volume expansion}
\label{latticeexp}
\begin{figure}
\includegraphics[scale=.9]{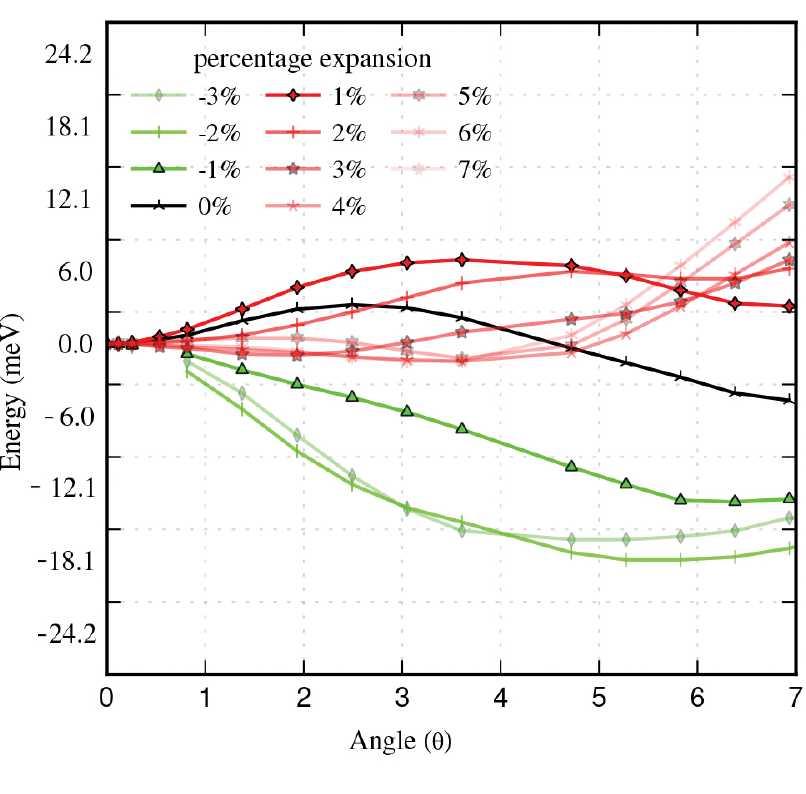}

\caption{Relative energy as function of  rotation angle at various lattice expansions for CsSnI$_3$ as \% expansions in lattice constant
  The energies are considered relative to the un-rotated energy at each lattice parameter \label{figrotation2}}
\end{figure}

As we already mentioned, the tendency toward octahedral rotation in the Sn and Pb halides
decreases, that is to say the rotation angle decreases, with increasing volume for a given material.
We therefore further studied the behavior under lattice expansion and compression, which one might
think of as occurring by thermal expansion and under high-pressure respectively.
First, we show the energy curves for CsSnI$_3$  as function of rotation angle for various
lattice expansions in Fig. \ref{figrotation2}. Even without volume expansion, we see that the curves shows two local minima, one at zero angle and one at about 7$^\circ$. As we increase the volume, the local minimum corresponding to the finite rotation moves up in relative energy and eventually,beyond 3 \% expansion of the lattice constants, it disappears, at which point the curve becomes very flat. Although they still show a very shallow finite angle
minimum, we may essentially consider this as a sign that the rotation is no longer preferred. 
On the other hand, under compression, the local
minimum appears to shift toward smaller angle and becomes deeper relative to the unrotated structure.
\begin{figure}

  \includegraphics[scale=0.35]{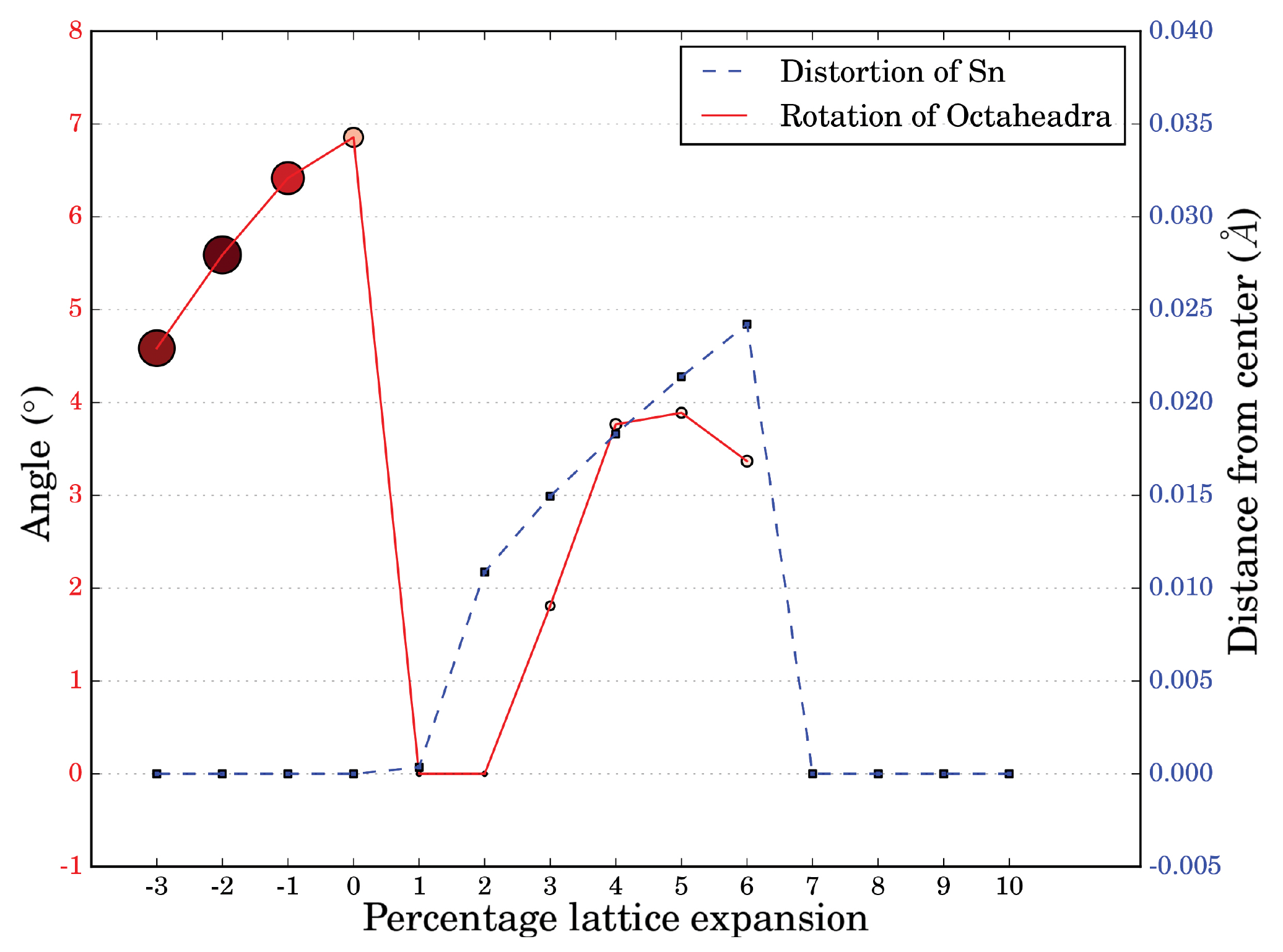}

  \caption{Percentage lattice-constant expansion {\em vs.} angle of rotation and distance of Sn from $[.5,.5,.5]$ for CsSnI$_3$. The size and darkness of the rotation markers (circular ones) represent the size of energy barrier $w.r.t$ the
    perfect cubic perovskite structure\label{figdistortexpansion}}
\end{figure}

The optimum angle of rotation is shown as the red curve as function of lattice  expansion in Fig. \ref{figdistortexpansion}. The increasing values for
lattice expansion actually correspond to a very low energy barrier,
as is indicated by the small sign of the symbols marking each point
and may to first approximation be ignored. Under compression, the rotation
angle clearly is reduced and the barrier increases, meaning the energy
of the rotated minimum becomes deeper. 

\begin{figure}

\includegraphics[scale=0.56]{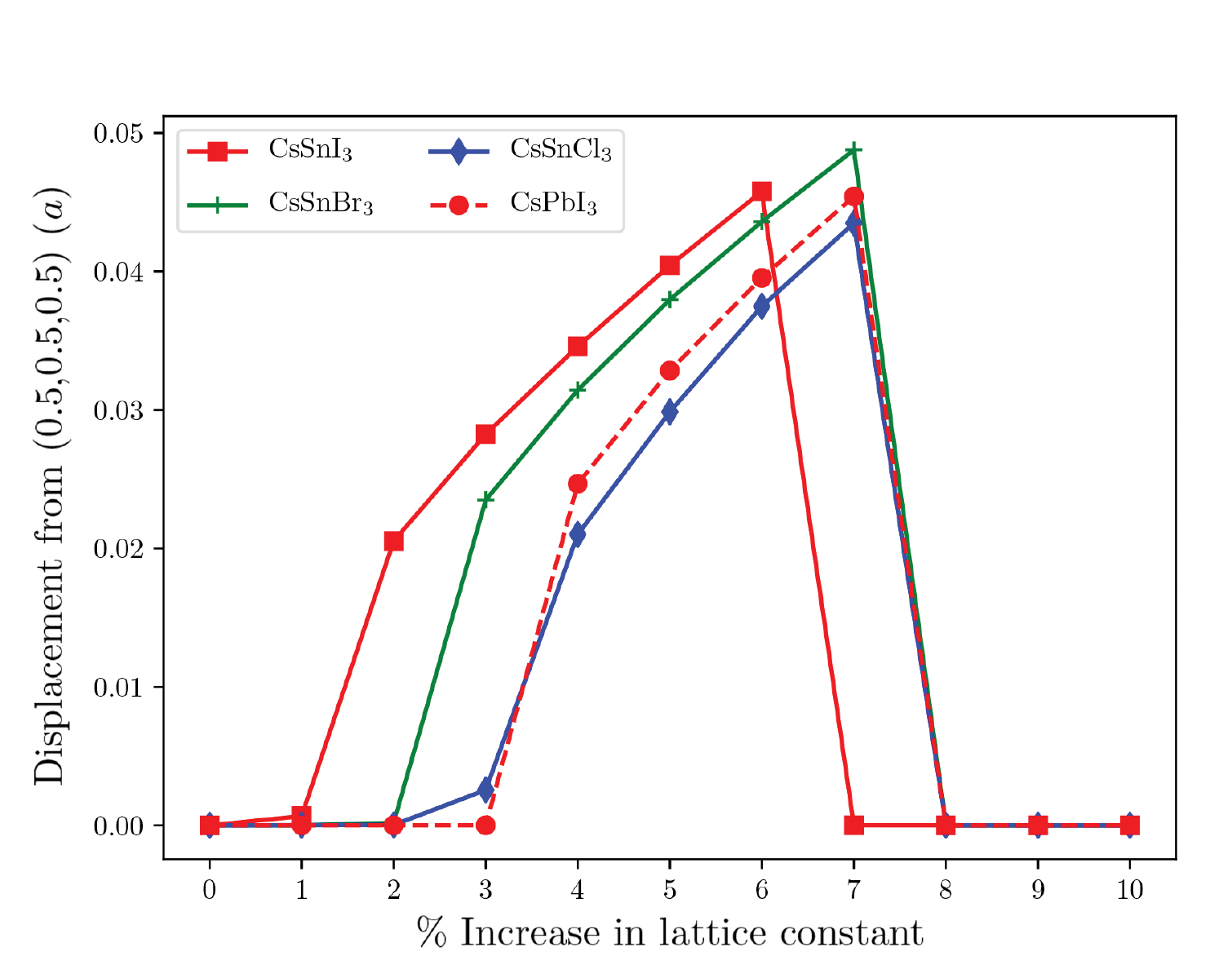}

\caption{Displacement of Sn/Pb from the center $[.5,.5,.5]$ for CsSnI$_3$, CsSnCl$_3$, CsSnBr$_3$ and CsPbI$_3$ as function of lattice constant expansion. \label{figdistort2}}
\end{figure}

Next, we examine the possibility of off-centering of the Sn atom as function of lattice expansion.
As we can see in Fig.  \ref{figdistortexpansion} in the blue dashed curve, the off-centering displacement stays zero
until 1 \% expansion at which point it starts increasing linearly. Eventually it collapses again beyond 6 \% expansion. 
Similar results are also obtained for the other halogens and for the Pb compounds as shown in Fig.  \ref{figdistort2}.
In summary, we find that beyond
a given lattice expansion the CsSnX$_3$ and CsPbX$_3$ materials undergo a rhombohedral distortion with off-centering
of the Sn (or Pb) rather than the octahedral rotation.

This type of behavior was reported earlier for CsSnBr$_3$ by Fabini \etal\cite{Fabini16} and related to the
active lone-pair behavior of the $s$ electrons which was studied in detail. 
We thus see that there is indeed a competition between the two types of distortion behavior, rotation
or rhombohedral off-centering.  The lone pair character promotes the off-centering and
is strongest for Ge and Si if the latter are required, as in this structure, to behave divalent
but it also occurs in Sn and to a smaller degree in Pb. However, in Sn and Pb this mechanism
of distortion is in competition with rotations, while in Ge and Si it is not.
Finally, it should be pointed out that the off-centering in CsSnBr$_3$ was experimentally observed
by Fabini \etal\cite{Fabini16} but occurs dynamically. It was observed only  through analysis of the pair distribution functions. 
In other words, it does not occur
coherently throughout the sample, which means that a rhombohedral crystallographic phase is not found for this compound.
Instead it is hidden in the cubic phase but is apparent from
the large atomic displacements which are coherent only on a local scale.   This is an important
difference from Ge where the rhombohedral phase is the actually observed equilibrium crystal structure.

\subsection{Rb instead of Cs}
\label{Rbcase}

\begin{table}[h] 
  \caption{Energy barrier and angle of rigidly rotated  RbGeX$_3$ with X=Cl, Br, I.The Energy barrier is the barriers between the cubic structure with 0$^\circ$ and local/global minimum at the given angle  \label{tabrotrbge}}
\begin{ruledtabular}
\begin{tabular}{lccc}
 Compound             & RbGeCl3       & RbGeBr3     & RbGeI3        \\ \hline\\
Angle ($\theta$)       & 4.71   & 6.93 & 10.21   \\
Energy barier (meV) & -70.0 & 13.5  & 36.4 \\ 
\end{tabular}
\end{ruledtabular}
\end{table}

In this section we consider the RbGeX$_3$ compounds  compared with the
CsGeX$_3$ compounds. From Table \ref{tabtolerance} we expect that
because of the smaller size of the Rb ion, these compounds would be unstable
toward octahedral rotation. The results for rotation in Table \ref{tabrotrbge}
show indeed that octahedral rotation lowers the energy for a finite
rotation angle for I and Br but not for Cl. In the latter case, there is
still a local minimum at a finite angle but its energy is actually higher than
at the zero angle rotation.  The energy lowering is comparable and even larger
than for the corresponding CsSnX$_3$ compounds and
the angle of rotation is larger for I than for Br.

\begin{table}[h] 
  \caption{Optimized cubic and
    rhomobohedral structures for RbGeX$_3$ with X=Cl, Br, I.
    The $\Delta E$ are the barriers between the cubic structure
    with $\delta u=0$ and $\eta=1$ and the optimized rhombohedral structure
    each at their own equilibrim volume. \label{tabrbgerhom}}
\begin{ruledtabular}
\begin{tabular}{llll}
Compound       & RbGeCl3       & RbGeBr3      & RbGeI3        \\ \hline \\
Cubic $a$ (\AA) GGA      & 5.345         & 5.57         & 5.97          \\
Cubic $V$ (\AA$^3$)      & 152.70   & 172.80   & 212.77 \\
Cubic bond length (\AA)      & 2.67&  2.78& 2.98 \\
Rhombohedral $a$ (\AA$^3$)   & 5.44      & 5.65     & 5.99      \\
Rhombohedral $V$ (\AA$^3$)   & 161.27 &180.44 &214.98      \\
Rhombohedral bond length (\AA)      & 2.31 &  2.46 &  2.69 \\
$\Delta V/V$ (\%) &5.31\%       & 4.23\%       & 1.02\%        \\
Change in bond length (\%) &-13.37\%       & -11.55\%       & -9.84\%        \\
$\delta u$         & 0.035      & 0.038     & 0.036     \\
$\delta_1$         & 0.001      & 0.002     & 0.001      \\
$\delta_2$         & 0.011      & 0.025     & 0.014      \\
$\eta$             & 1.026      & 1.052     & 1.039      \\
$\alpha$          & 88.47   & 87.00  & 87.71   \\
$\Delta E$ (meV) GGA    & 455.8 & 367.2 & 304.7\\
Band Gap (eV) GGA    & 1.72 & 1.05  & 0.43
\end{tabular}
\end{ruledtabular}
\end{table}

On the other hand, from the previous sections, it is also clear that from
the point of view of lone-pair physics, Ge is prone to off-centering.
Therefore we also study the possibility of lowering the energy by the
rhombohedral disotortion. The results are shown in Table
\ref{tabrbgerhom}.
This shows that the off-centering and related rhombohedral distortion lowers
the energy significantly more efficiently than the octahedral rotation.
For RbGeCl$_3$ the rotation actually does not lower the energy, while
the distortion does. For RbGeBr$_3$ and RbGeI$_3$, the energy lowering
by the off-centering is significatnly larger than by rotation of the octahedra.
Thus comparing
to the Cs case, this indicates that the off-centering of Ge is not so much determined
by the tolerance factor but rather by the lone-pair physics.
The relaxation parameters, barriers and energy gaps in GGA for these compounds are given
in Table \ref{tabrbgerhom} in the same way as for the other compounds.

\begin{figure}
  \includegraphics[width=8cm]{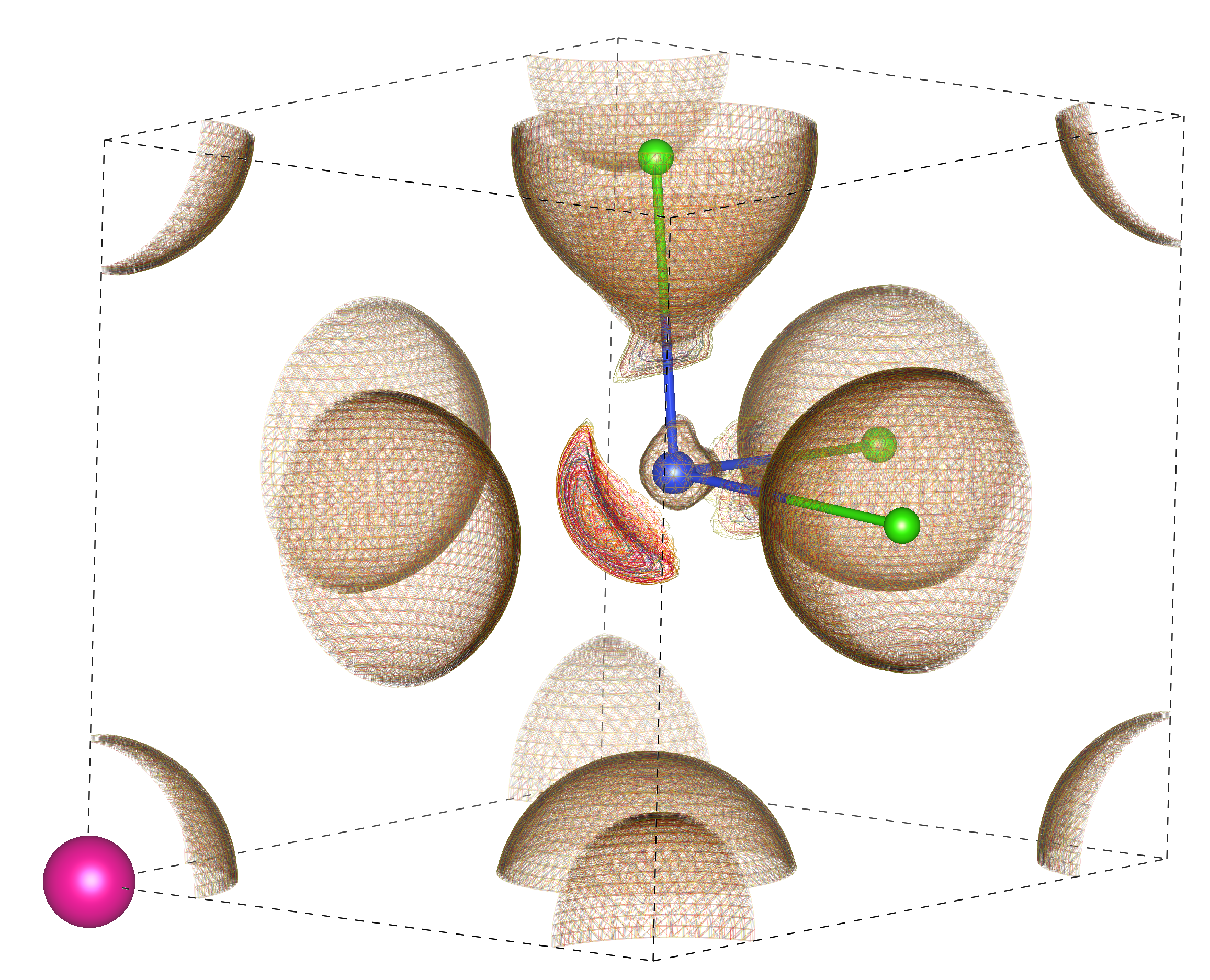}
  \caption{Total valence charge density for the relaxed distorted RbGeCl$_3$ shown as a superposition of 8 isosurfaces with values ranging frm 0.058
    to 0.071 $e/a_0^3$. Each isosorface is shown as a mesh
    of different color. One can clearly distinguish the Ge-$s$ like lobe
    in the direction opposite to the displacemnt.
    The pink sphere is Rb, the blue one Ge and the green ones Cl.\label{figlone3d}}
\end{figure}

\begin{figure}
  \includegraphics[width=8 cm]{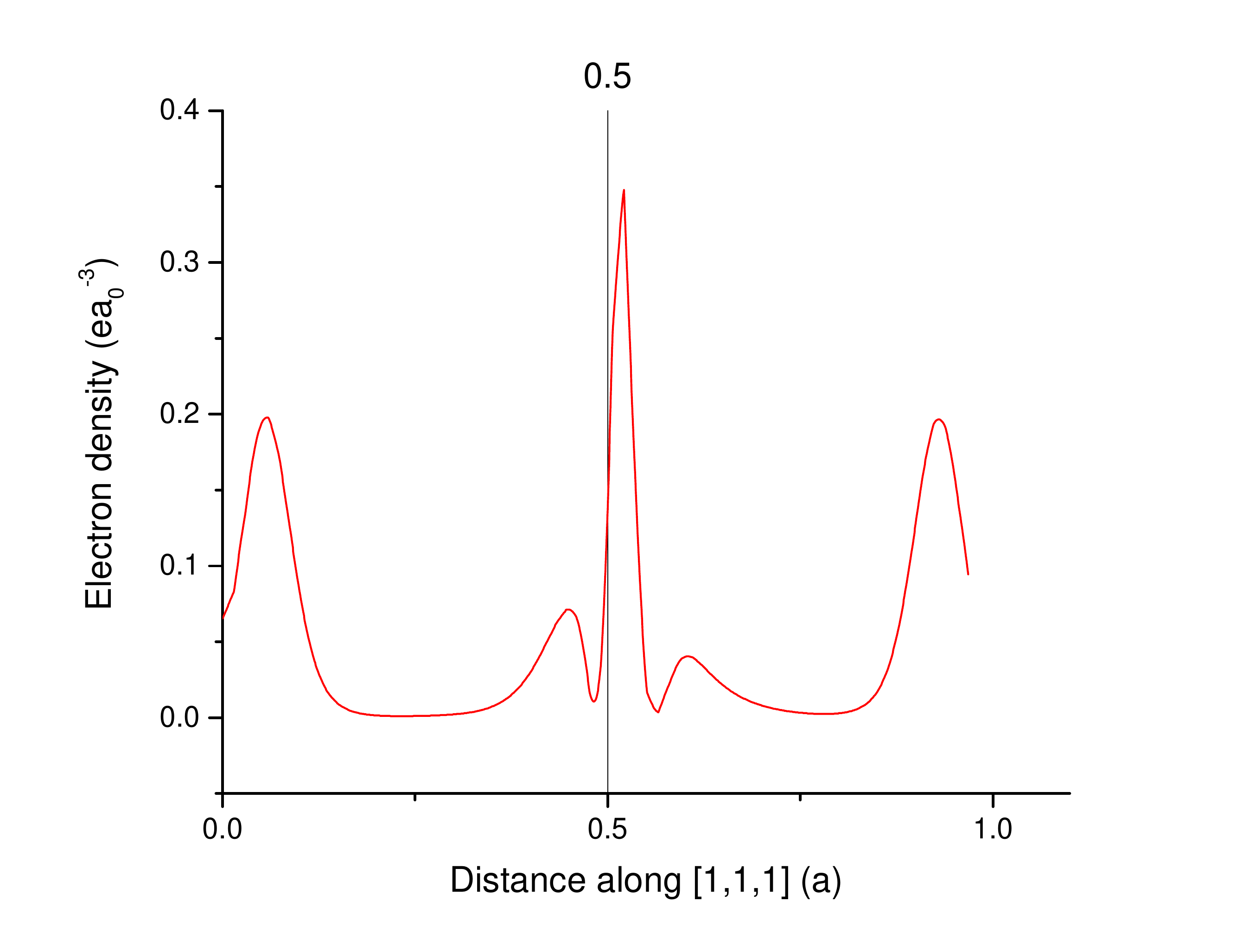}
  \caption{Plot of the valence charge density along the [111] body diagonal.
    One can see the asymmetry of the charge density near the Ge position,
    at a position larger than 0.5, again
  reflecting the lone-pair character.\label{figlonediag}}
\end{figure}

Finally, we illustrate the lone-pair character in this case by plotting the
charge density for this in Figs. \ref{figlone3d}, \ref{figlonediag}.
The firts one shows a 3D view of isosurfaces, the second one shows the
valence charge density along the body diagonal. 

\section{Conclusions} \label{conclusions}
In this paper we have explored the stability of inorganic halide perovskites ABX$_3$
with $X$ a halogen (Cl, Br, I), $A$ a large alkali ion(Cs or Rb) and B a group IV element,
(Si, Ge, Sn, Pb), under two types of
distortion: an antiferroelectric distortion corresponding to octahedral rotation  and a ferro-electric
off-centering of the central IV ion inside its halogen octahedron.
At first, we find that there is a clear trend that the Pb and Sn cases prefer rotation while
Ge and Si prefer ferro-electric distortion. We also find that the rotation, when fully optimizing
the structures, is accompanied by a reduction of the volume. The off-centering is accompanied by
rhombohedral distortion and volume increase.  The tendency toward rotation is clearly related to the
Goldschmidt tolerance factor. On the other hand, we find that upon volume expansion, the
rotation angle decreases and beyond a certain expansion off-centering becomes favorable
even for Sn and Pb based compounds. The origin of the off-centering is thus more related to the
lone-pair physics. The Ge and Si based compounds, in which Ge and or Si are forced to behave as a
divalent ion, strongly favor lone-pair induced off-centering or ferro-electric distortion.
In the Rb case, both distortion modes tend to lower the energy (except for the Cl case)
but the ferro-electric distortion nonetheless lowers the energy signiricantly more efficiently.
Thus the lone-pair physics dominates the RbGeX$_3$ based compounds rather than the tolerance factor related rotation. 
The two distortion mechanisms can thus be in competition with each other
and the off-centering for the Si and Ge cases occurs even if the
tolerance factor would allow for rotations as a mechanism to lower the energy.

\acknowledgments{This work was supported by the U.S. Department of Energy (Basic Energy Sciences)
  DOE-BES under grant No. DE-SC0008933. The calculations made use of the High Performance Computing Resource in the Core Facility for Advanced Research Computing at Case Western Reserve University.}

\bibliography{dft,csnx,gw,lmto}
\end{document}